# Static and Dynamic Analyses of Free-Hinged-Hinged-Hinged-Free Beam in Non-Homogeneous Gravitational Field: Application to Gravity Gradiometry

Alexey V. Veryaskin[1,2] and Thomas J. Meyer[3]

[1]Trinity Research Labs, School of Physics, Mathematics and Computing, University of Western Australia, 35 Stirling Highway, Crawley, WA 6009, Australia
[2]Quantum Technologies and Dark Matter Research Laboratory (QDM Lab), Department of Physics, University of Western Australia, 35 Stirling Highway, Crawley, WA 6009, Australia
[3]Lockheed Martin RMS – Gravity Systems, 2221 Niagara Falls Boulevard, Niagara Falls, NY 14304.
Email: tom.j.meyer@lmco.com ; alexey.veryaskin@uwa.edu.au

**Summary**

The first analytical evaluation of free-hinged-hinged-hinged-free beam, proposed to be used as a primary sensing element for gravity gradiometry, is presented. The results of the evaluation, obtained in quadratures, are applied to the beam's structure such as the locations of hinges that form the beam's boundary conditions allowing only free rotations around the nodal axes. The latter are purposely chosen to minimize the beam's symmetric free ends deflection under the uniform force of gravity while simultaneously permitting the beam's maximum possible mirror-symmetric free ends deflection due to a gravity gradient along its length. The flexible triple hinged beam's deflection from its unperturbed position is internally entangled at all locations including free ends and this allows for synchronized mechanical displacement measurements at any beam's deflection point. Some methods of manufacturing such sensing element and the corresponding error factors are also discussed and presented for the first time.

**Introduction**

Distributed flexible mechanical objects such as cantilevers[1], flexures used in MEMS[2,3,4] and strings/ribbons[5] have been in use for quite a long time as primary sensing elements (test masses) for measuring gravitational acceleration and its spatial derivatives (gravity gradients). By measuring the latter, one could get, for example, valuable information about buried mineral deposits, hidden underground voids, tunnels and bunkers, and use the fine Earth's gravity data for passive, not jammable, strategic submarine navigation[6]. The difference between the primary sensing elements in the form of distributed flexible test masses (elongated beams and ribbons) and compact solid test masses (spheres, cylinders and the like) is that in the latter case the test masses are responsive to the local force of gravity applied to centres-of-mass, while in the former case they can be more receptive to the force-per-unit-length that is, by definition, directly proportional to gravity gradients. The distributed test masses allow for the construction of a continuous beam type gravity gradiometer where only one sensing element is needed for measuring a gravity gradient along its length[7]. If compact test masses are used, a minimum of two of them are needed to measure the difference in the force of gravity acting upon them. In the latter case one needs to maintain a fixed spatial separation of the test masses or "baseline" which effectively determines sensitivity and size of the resulting gradiometer instrument. The stand-alone test masses and the corresponding sensing means, that translate their motion into a measurable physical quantity, must be matched with an unprecedented accuracy in order to meet a state-of-the-art measurement capability (typically within a few parts per billion for requisite ultra-precision) which, in turn, must be provided over the whole gradiometer run time, i.e. must be stable within reasonably long time intervals. A primary sensing element having a single sensitivity axis is shown in Fig.1 below. It comprises an elongated thin metal beam (ribbon or foil) which is hinged at three equally separated axes and having overhanging free ends[8]. In beam theory it might be called a free-hinged-hinged-hinged-free beam. The hinges are connected to so-called zero-force frame of reference providing free beam's rotational motion around their axes.

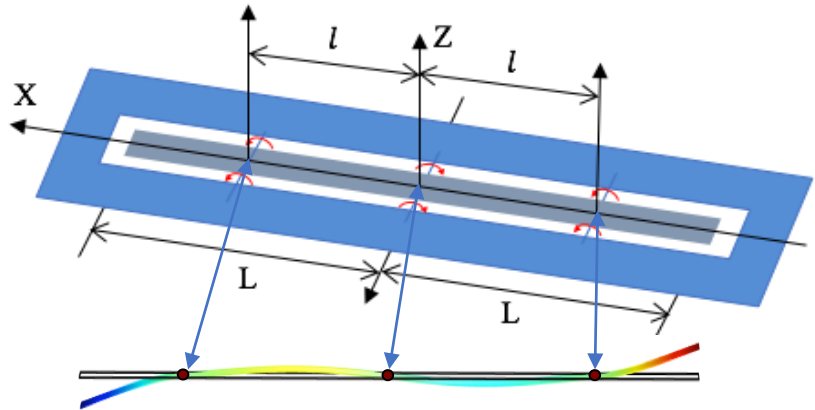

**Figure 1**. A mirror-symmetric profile of an elongated thin metal beam (the gray colored ribbon in the middle of a solid blue reference frame) which is hinged at three equally separated axes and having overhanging free ends. Such deformation profile also shown is the result of a linearly distributed force-per-unit-length, representing a pure gravity gradient load.

The deflection of the beam's free ends with respect to its unperturbed position can be measured by an appropriate mechanical-displacement-to-voltage conversion technique such as a moving plate capacitive transducer[9,10]. The locations of the hinges are sought to simultaneously minimize the beam's symmetric, with respect to the central axis, free ends deflection under the uniform force distribution while permitting the beam's maximum possible mirror-symmetric deflection due to a gravity gradient along its length. The beam's free-hinged-hinged-hinged-free structure is heavily over-constrained, and to the best of the authors' knowledge has not been analytically evaluated before in open publications. For the first time a static and dynamic analyses, including error factors, of such over-constrained beam loaded by a non-uniform force distribution is presented below. This approach is entirely new in gravity gradiometry and may open the door to the most desired use of gravity gradiometers in the strapped-down mode onboard commercial drones and other unmanned platforms.

**Theoretical Framework**

There are well known theoretical frameworks that have been widely used to analyse the general transverse motion of elongated flexible beams under different boundary conditions, which are Euler-Bernoulli-Lagrange theory[11] and Rayleigh-Timoshenko theory[12,13,14]. In the static approximation, they coalesce to the same framework where the transverse displacement $Z$ of every point of a beam under investigation and its bending slope $\theta$ are described by the following equations

$$\frac{EI}{\eta}\frac{d^4Z}{dx^4} = g_z(x,z) \cong g_z(0) + \Gamma_{zx}x \tag{1}$$

$$\frac{EI}{\eta}\frac{d^4\theta}{dx^4} = \frac{d}{dx}g_z(x,z) \cong \Gamma_{zx} \tag{2}$$

$$\theta = \frac{dZ}{dx} \tag{3}$$

where $E$ is the beam's Young modulus of elasticity, $I$ is the beam's area moment of inertia, $\eta$ is the beam's mass per unit length, $g_z$ is normal to the beam surface gravitational acceleration vector component and $\Gamma_{zx}$ is a gravity gradient tensor component along the beam's length

$$\Gamma_{zx} = \frac{g_z(L) - g_z(-L)}{2L} \tag{4}$$

The beam's area moment of inertia $I$ is described by the following equation[15,16]

$$I = \frac{1}{12}\frac{bd^3}{1-\sigma^2} \tag{5}$$

where $b$ and $d$ are the width (base) and thickness of the beam accordingly, $\sigma$ is Poisson ratio.

In Eq.(1) and Eq.(2) an extra term $\Gamma_{zz}z$ has been ignored in the first order series expansion of the gravitational acceleration component $g_z(x,z)$ over coordinates $x$ and $z$, where $\Gamma_{zz}$ is the vertical gravity gradient component. This term, if left there, would introduce so-called gravitational spring, that can be either positive or negative, since $Z=z$ represents the beam's transverse displacement variable. Including this term leads to unnecessary complication of the quasi-static analysis as the corresponding corrections are the second order of magnitude. Also, this term is a symmetric one with respect to $x \to -x$ coordinate transfer and therefore can not modify the beam's deflection under the gradient term $\Gamma_{zx}$.

For the free-hinged-hinged-hinged-free beam the following boundary conditions are applied to Eq.(1) and Eq.(2)

$$Z(-l) = Z(0) = Z(l) = 0 \tag{6}$$

$$\frac{d^2\theta}{dx^2}\bigg|_{x=-L} = \frac{d^2\theta}{dx^2}\bigg|_{x=L} = 0 \tag{7}$$

$$\frac{d\theta}{dx}\bigg|_{x=-l} + \frac{d\theta}{dx}\bigg|_{x=l} + 4\frac{d\theta}{dx}\bigg|_{x=0} = \frac{1}{2}\eta l^2 g_z(0) \tag{8}$$

Eq.(8) above is the manifestation of the "three moment" theorem that is applicable to any flexible beam's adjoint three-span sections[8]. In the error-free model above, it is assumed that the hinges are aligned along the same XOY plane and separated by exactly the same spans. Also, the beam's free ends are separated by exactly the same distance from the middle point x=0 of the static coordinate system XYZ where the X-axis coincides with the unperturbed position of the beam. It is also assumed that the beam's length (2L) is much larger compared to its width (b) and thickness (h) so the one-dimensional problem above is well-justified[17].

The Eq.(1) and Eq.(2) cannot be solved in quadratures for the whole length of the beam due to the number of boundary conditions vastly exceeding the order of the linear differential equations. However, they can be solved for each of the four spans along the beam's length provided the continuity of the solutions at the nodal points is preserved[18]

$$\theta(l + \varepsilon) = \theta(l - \varepsilon), \varepsilon \to 0 \tag{9}$$

$$\theta(-l - \varepsilon) = \theta(-l + \epsilon), \epsilon \to 0 \tag{10}$$

$$\theta(0 + \varepsilon) = \theta(0 - \varepsilon), \epsilon \to 0 \tag{11}$$

$$\frac{d\theta}{dx}|_{l+\epsilon} = \frac{d\theta}{dx}|_{l-\varepsilon}, \varepsilon \to 0 \tag{12}$$

$$\frac{d\theta}{dx}|_{-l-\epsilon} = \frac{d\theta}{dx}|_{-l+\varepsilon}, \varepsilon \to 0 \tag{13}$$

$$\frac{d\theta}{dx}|_{0+\epsilon} = \frac{d\theta}{dx}|_{0-\varepsilon}, \varepsilon \to 0 \tag{14}$$

It is worth noting that the bending moments $EId\theta/dx$ at the beam's free ends do not vanish in the presence of the force gradient along the beam's length. For the sake of simplicity, the details of solving Eq.(1) and Eq.(2) for four adjoint spans independently are omitted here and the final results are shown below

$l \leq x \leq L$

$$Z(x) = \\ = -\Gamma_{zx}\eta \frac{(l-x)\left(10l^4 - 27l^3 x - 30L^2 x^2 + 3x^4 - l^2(30L^2 - 3x^2) + 3l(20L^2 x + x^3)\right)}{360EI} \\ + g_z(0)\eta \frac{(l-x)\left(l^3 - 2l^2(2L+x) + l(6L^2 + 8Lx - 2x^2) - 2x(6L^2 - 4Lx + x^2)\right)}{48EI} \tag{15}$$

$$\theta(x) = \Gamma_{zx}\eta \frac{37l^4 - 90l^2 L^2 - 60l^3 x + 180lL^2 x - 15x^2(6L^2 - x^2)}{360EI} \\ - g_z(0)\eta \frac{3l^3 - 12l^2 L + 18lL^2 - 8x(3L^2 - 3Lx + x^2)}{48EI} \tag{16}$$

$0 \leq x \leq l$

$$Z(x) = \Gamma_{zx}\eta \frac{x(7l^4 - 10l^2 x^2 + 3x^4)}{360EI} - g_z(0)\eta \frac{(l-x)x^2\left(3l^2 + 6L^2 - 2l(6L - x)\right)}{48EIl} \tag{17}$$

$$\theta(x) = \Gamma_{zx}\eta \frac{7l^4 - 30l^2 x^2 + 15x^4}{360EI} \\ - g_z(0)\eta \frac{x\left(6l^3 - 18L^2 x - 3l^2(8L+x) + 4l(3L^2 + 9Lx - 2x^2)\right)}{48EIl} \tag{18}$$

$-L \leq x \leq -l$

$$Z(x) =$$
$$= \Gamma_{zx}\eta \frac{(l+x)(10l^4 + 27l^3x - 30L^2x^2 + 3x^4 - l^2(30L^2 - 3x^2) - 3l(20L^2x + x^3))}{360EI}$$
$$+ g_z(0)\eta \frac{(l+x)(l^3 - 2l^2(2L-x) + l(6L^2 - 8Lx - 2x^2) + 2x(6L^2 + 4Lx + x^2))}{48EI} \quad (19)$$

$$\theta(x) = \Gamma_{zx}\eta \frac{37l^4 - 90l^2L^2 + 60l^3x - 180lL^2x - 15x^2(6L^2 - x^2)}{360EI}$$
$$+ g_z(0)\eta \frac{3l^3 - 12l^2L + 18lL^2 + 8x(3L^2 + 3Lx + x^2)}{48EI} \quad (20)$$

$-l \le x \le 0$

$$Z(\mathrm{x}) = \Gamma_{zx}\eta \frac{x(7l^4 - 10l^2x^2 + 3x^4)}{360EI} - g_z(0)\eta \frac{(l+x)x^2(3l^2 + 6L^2 - 2l(6L+x))}{48EIl} \quad (21)$$

$$\theta(x) = \Gamma_{zx}\eta \frac{7l^4 - 30l^2x^2 + 15x^4}{360EI}$$
$$- g_z(0)\eta \frac{x(6l^3 + 18L^2x - 3l^2(8L - x) + 4l(3L^2 - 9Lx - 2x^2))}{48EIl} \quad (22)$$

One can find the condition upon which the beam's displacements at its free ends under the uniform force of gravity vanish ($L \ne l$). One has

$$Z(L) = g_z(0)\eta \frac{(l-L)(l^3 - 6l^2L + 12lL^2 - 6L^3)}{48EI} = 0 \quad (23)$$

It yields

$$l = (2 - 2^{1/3})L \quad (24)$$

It is interestingly enough to note that the positions $\pm l$ in Eq.(24) match closely with free-free beam's nodal locations of the first mirror-symmetric eigenmode as shown in Fig.2 below[19]

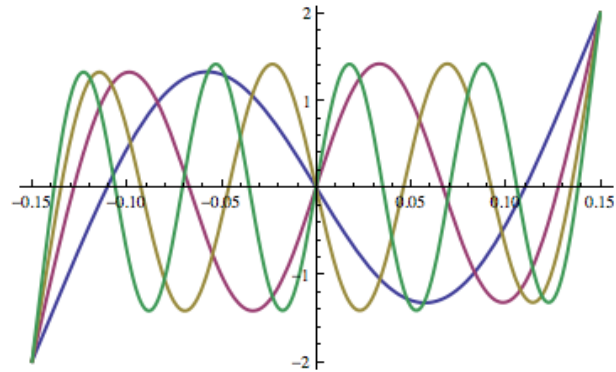

**Figure 2**. Mirror-symmetric eigenmodes of a free-free 30cm thin beam. The first mirror-symmetric eigenmode is depicted in blue. The eigenmode's zero-crossing locations are at 0, ±11.0368 cm compared to 0, ±11.1012 cm derived from Eq.(24) for the free-hinged-hinged-hinged-free beam of the same length.

The deflections of the free-hinged-hinged-hinged-free beam's ends with the locations of the hinges as per Eq.(24) are as follows

$$Z(L) = -Z(-L) = -2.2 \times 10^{-3} \Gamma_{zx} \eta \frac{L^5}{EI} \cong -10^{-3} \Gamma_{zx} \frac{mL^4}{EI} \qquad (25)$$

where *m* is the total mass of the beam and *mL⁴/EI* can be treated as the beam's intrinsic gradiometric gain having the dimension of the product of time squared and distance. It is also interesting to note that this parameter can be measured quite accurately for particular dimensions and material. As an example, the product *EI* can be extracted from measuring the resonant frequencies (say, the first resonant mode) of a standard cantilever fixed-free beam of the same material and cross section. In turn, this means that absolute gravity gradient measurements are possible provided that the deflection of the beam from its force-free position can also be measured in absolute units.

In Fig.3 below, two beam's spatial profiles show its deflection under full-g body load and, independently, under 10E gravity gradient along its length (g = - 9.8 m/s² is the Earth's gravitational acceleration chosen to be directed downwards, 1E = 1Eotvos = $10^{-9}$ 1/s² is the unit of gravity gradients, *b=0.01metres, d=0.00025metres, l = ±(2 − $2^{1/3}$)L*).

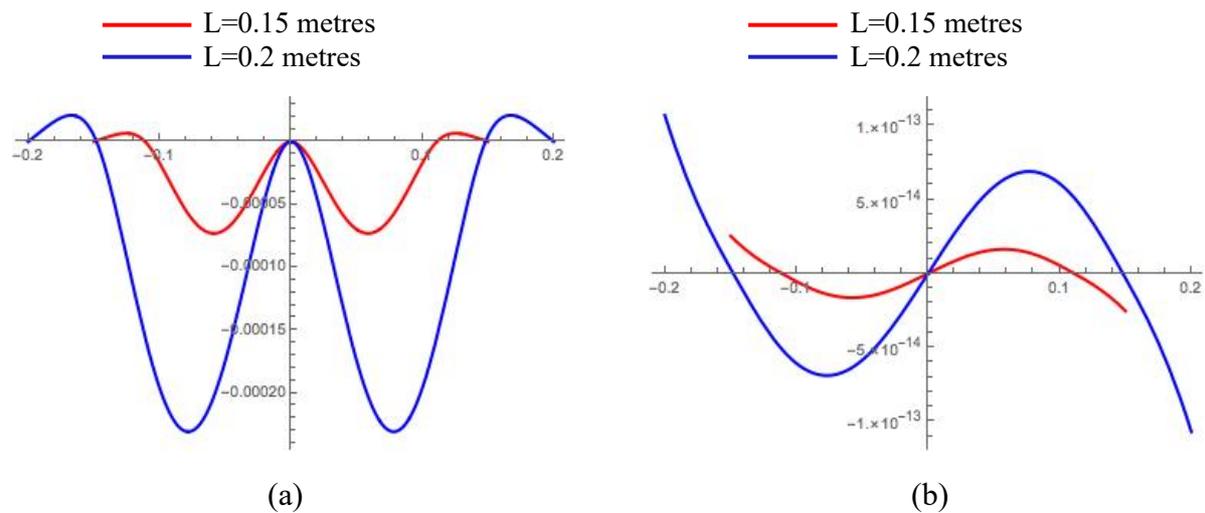

(a)　　　　　　　　　　　　　　　　　(b)

**Figure 3**. (a) – the symmetric deflection of the free-hinged-hinged-hinged-free beam under the full projection of the Earth's gravity; (b) – the mirror-symmetric deflection of the beam under 10E gravity gradient (all numbers are in metres, material is Phosphor Bronze).

**Design and Manufacturing Error Analysis (static)**

Manufacturing processes are not perfect, and the ideal scenario outlined above can be modified to account for several imperfections, including:

- Asymmetric positioning of the left-right side hinges with respect to the central one, so the beam's spans are not equal to each other;

- The beam's mass per unit length is not the same at every cross-section of the beam (major contributions are beam's manufacturing tolerances and its density variations);

One can prove that all such effects result in mixing the true gravity gradient related deflection of the beam with the one caused by the uniform force of gravity distribution. As an example, the deflection of the beam's right end, including an error term caused by the mispositioning of the right-side hinge $\ell + \delta\ell$ and asymmetric right end $L + \delta L$, is as follows

$$Z(L) \cong -10^{-3}\Gamma_{zx}\frac{mL^4}{EI} + g_z(0)\frac{mL^3}{96EI}\left(\alpha\frac{\delta\ell}{L} + \beta\frac{\delta L}{L}\right) + Z_0(L) \qquad (26)$$

where $\alpha$ and $\beta$ are numerical factors of the order of unity and $Z_0$ is a zero-force offset due to a residual stress embedded into the beam. In real life, there always will be some small residual curvature of the beam's plane due to the remaining (built-in) stress embedded into its atomic structure and caused by its history of mechanical and thermal treatment during manufacturing.

If the stress is constant (frozen) in time, then this appears as the presence of a constant uniform load or a constant gravity gradient along the ribbon's length independent of the environmental conditions. These will result in a systematic error in measuring gravity gradients in absolute units, namely, a measurement bias – a difficult situation. As with most sensitive equipment, relative measurements are much easier to deal with provided the relevant gradiometer's set-up is stable – the requirement that makes the development of practical gravity gradiometers look similar to "impossible" grade missions. However, in the case of such spatially distributed flexible sensing element, i.e. a free-hinged-hinged-hinged-free beam, the biased deflections of its different spans can be measured independently and combined (in real time or in a post-processing stage) in such a way that this would cancel out the systematic errors mentioned above. This would also cancel out such dynamic effects as mixing desired signals with linear acceleration if a gradiometer is mounted on a moving platform (airborne gravity gradiometry as an example). The latter is possible since the forced mechanical displacements (either resonant or non-resonant) of every infinitesimally small cross section of vibrating beams are superposed as a weighted sum of all possible eigenmodes, formed by beam's boundary conditions, and representing true standing waves[20].

The design and manufacture of free-hinged-hinged-hinged-free beams is a challenging engineering problem. The analysis above assumes the hinges are solid structures that do not possess any intrinsic torsional spring constant. Such hinges can be made as solid micro-shafts allowing only free rotation of the beam around the nodal axes and connected to an external frame of reference in a manner that used in the precision wrist-watch making technique, e.g. certified mechanical chronometers. The latter allows for a self-alignment of the beam's nodal axes along the zero-force plane (XOY in the case above). Such technique can also mitigate a residual mismatch of the thermal expansion of the materials used to make a beam, shafts and their holders and the frame of reference. All such materials must have very closely matched thermal expansion coefficients. An example of a solid shaft-based design of a free-hinged-hinged-hinged-free beam is depicted in Fig.4 below.

Another design of a free-hinged-hinged-hinged-free beam is depicted in Fig.5 below. The whole structure is EDM wire cut with about 3-5 microns tolerance from a single piece of flat metal foil. In this case, the hinges represent micro-pivots of an optimum cross section connecting the middle flexible section (ribbon) to the rest of the foil that form a zero-force frame of reference. In turn, the latter can be firmly mounted upon another solid frame made of the same material as the foil. The pivots possess intrinsic torsional spring constant and this changes the boundary conditions of the true hinged beam design[21,22]. Analytical evaluation becomes complicated as the additional torsional rigidity is not well known and must therefore be idealized. During the EDM processing time, the fine pivot structure experiences not well-defined treatment by heating the foil material in a water basin, which is the standard processing environment for the EDM technique. The effect of the latter upon a particular chosen material is not well known either. The cutting wire material does matter as well. Software modelling is needed to simulate these effects and compare the results with experimental data. A detailed error analysis of a prototype gravity gradiometer along with experimental data will be published elsewhere.

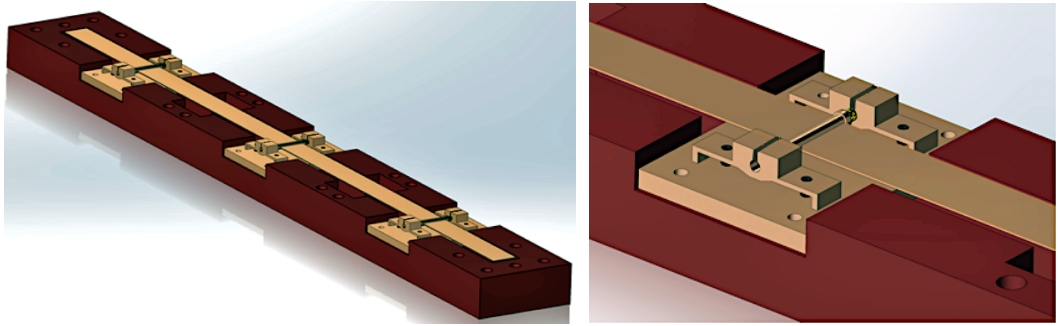

**Figure 4**. A free-hinged-hinged-hinged-free beam made by sliding a straight long ribbon into ultra-small and almost weightless shafts manufactured by experienced wrist-watch makers [http://www.wcawa.org.au]. The shafts are locked from both sides and held inside jewels-bearing shaft holders fixed at nodal positions along the beam's length within a few micro-metres tolerance.

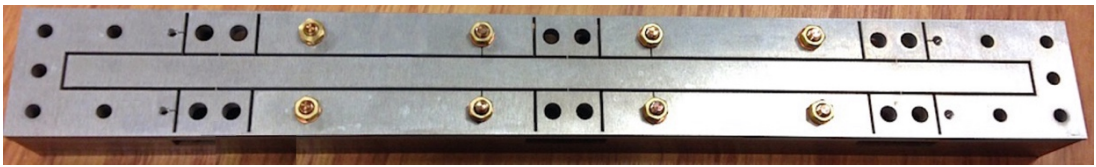

**Figure 5**. A free-hinged-hinged-hinged-free beam made by precision EDM process with about 3-5 microns tolerance. The hinges represent micro-pivots connecting the middle flexible section (ribbon) to the rest of the frame. Material shown is a composite Tungsten/Copper alloy.

**Dynamic Analysis of Free-Hinged-Hinged-Hinged-Free Beam**

The dynamic analysis of free-hinged-hinged-hinged-free beam is different for the Rayleigh-Timoshenko theory and for the Euler-Bernoulli-Lagrange theory. Timoshenko introduced corrections to the latter taking into account rotary inertia and shear deformation which leads to two linear differential equations for beam's transverse deflection and its bending slope[23]. This gives better results in dynamic analyses for beam's high frequency eigenmodes. However, for the multi-span thin beam under consideration the latter are vastly suppressed and only a few low frequency eigenmodes do matter for practical applications. Assuming the Euler-Bernoulli-Lagrange theory still accurate enough to analyse the dynamic behaviour of the beam under consideration[24], one has the following dynamic Euler-Lagrange equations for the beam's transverse displacement and bending slope

$$\eta \frac{\partial^2 Z}{\partial t^2} + h \frac{\partial Z}{\partial t} + EI \frac{\partial^4 Z}{\partial x^4} = -\eta a(t) + \eta \tilde{\Gamma}_{zx} x + F_L(x,t) \qquad (27)$$

$$\eta \frac{\partial^2 \theta}{\partial t^2} + h \frac{\partial \theta}{\partial t} + EI \frac{\partial^4 \theta}{\partial x^4} = \eta \tilde{\Gamma}_{zx} + \frac{\partial}{\partial x} F_L(x,t) \qquad (28)$$

$$\tilde{\Gamma}_{zx} = -\frac{d\Omega_y}{dt} - \Omega_z \Omega_x \qquad (29)$$

where $h$ is the coefficient of friction per unit length (assumed to be the same along the beam's length), $F_L(x,t)$ is the effective Langevin force *per unit length* forcing the beam to stay in thermal equilibrium with external environment with temperature $T$, $a(t)$ and $\Omega_{X,Y,Z}$ are kinematic (linear) acceleration normal to the beam's surface and angular velocity vector components of the beam's reference frame accordingly. The term $\tilde{\Gamma}_{zx}$ in the right side of Eq.27 and Eq.28 is called a dynamic gradient that affects the measurement of real gravity gradients by gravity gradiometers mounted on moving platforms representing non-inertial reference frames[25].

As the beam represents a multi-mode mechanical resonator, its response to the noise driving force is not the same as that of a Brownian particle. The intensity of the effective Langevin force per unit length should be calculated from the condition that the mean energy for each mode $n$ of the resonator will be given by $<W_n> = k_B T$[26]. A wave-function analysis applied to the free-hinged-hinged-hinged-free beam is presented below (see also[27,28]). One has

$$Z(x,t) = \sum_n \alpha_n(t) \psi_n^{(+)}(x) + \sum_n \beta_n(t) \psi_n^{(-)}(x) \qquad (30)$$

$$\int_{-L}^{L} dx\, \psi_n^{(+)}(x)\psi_m^{(+)}(x) = 2L\delta_{mn}, \quad \int_{-L}^{L} dx\, \psi_n^{(-)}(x)\psi_m^{(-)}(x) = 2L\delta_{mn} \qquad (31)$$

$$\int_{-L}^{L} dx\, \psi_n^{(+)}(x)\psi_m^{(-)}(x) = 0, \qquad \delta_{mn} = 0\ m \neq n, \qquad \delta_{nn} = 1 \qquad (32)$$

The modal wave-functions $\psi_n^{(+)}(x)$ and $\psi_n^{(-)}(x)$ represent symmetric and mirror-symmetric eigenfunctions satisfying the following relations

$$\psi_n^{(+)}(x) = \psi_n^{(+)}(-x), \qquad \psi_n^{(-)}(x) = -\psi_n^{(-)}(-x) \qquad (33)$$

These eigenfunctions are found by solving the following characteristic equation for each of the beam's span

$$\frac{d^4}{dx^4} \psi_n^{(\pm)}(x) = k_{(\pm),n}^4\, \psi_n^{(\pm)}(x) \qquad (34)$$

where $k_{(\pm),n}$ are the partial modal eigenvalues that depend on specific sets of boundary and continuity conditions corresponding to the beam's force-free vibration modes.

The orthogonality of the modal eigenfunctions is automatically provided by the beam's free ends boundary conditions (Eq.(35) and Eq.(36) below) and by their symmetry (Eq.(33))

$$\frac{d^2\psi_n^{(\pm)}}{dx^2}|_{x=-L} = \frac{d^2\psi_n^{(\pm)}}{dx^2}|_{x=L} = 0 \tag{35}$$

$$\frac{d^3\psi_n^{(\pm)}}{dx^3}|_{x=-L} = \frac{d^3\psi_n^{(\pm)}}{dx^3}|_{x=L} = 0 \tag{36}$$

The boundary and continuity conditions at $\pm\ell$ nodal locations for either symmetric or mirror-symmetric eigenfunctions are as follows

$$\psi_n^{(\pm)}(-l) = \psi_n^{(\pm)}(l) = 0 \tag{37}$$

$$\frac{d\psi_n^{(\pm)}}{dx}|_{l+\epsilon} = \frac{d\psi_n^{(\pm)}}{dx}|_{l-\epsilon}, \varepsilon \to 0 \tag{38}$$

$$\frac{d\psi_n^{(\pm)}}{dx}|_{-l-\epsilon} = \frac{d\psi_n^{(\pm)}}{dx}|_{-l+\epsilon}, \varepsilon \to 0 \tag{39}$$

$$\frac{d^2\psi_n^{(\pm)}}{dx^2}|_{l-\epsilon} = \frac{d^2\psi_n^{(\pm)}}{dx^2}|_{l+\epsilon} = 0, \varepsilon \to 0 \tag{40}$$

$$\frac{d^2\psi_n^{(\pm)}}{dx^2}|_{-l-\epsilon} = \frac{d^2\psi_n^{(\pm)}}{dx^2}|_{-l+\epsilon} = 0, \varepsilon \to 0 \tag{41}$$

For the central hinged location ($x=0$) one has

$$\psi_n^{(\pm)}(0) = \psi_n^{(\pm)}(0) = 0 \tag{42}$$

$$\frac{d\psi_n^{(+)}}{dx}|_{0-\epsilon} = -\frac{d\psi_n^{(+)}}{dx}|_{0+\varepsilon}, \varepsilon \to 0 \tag{43}$$

$$\frac{d^2\psi_n^{(+)}}{dx^2}|_{0-\epsilon} = \frac{d^2\psi_n^{(+)}}{dx^2}|_{0+\varepsilon} = 0, \varepsilon \to 0 \tag{44}$$

$$\frac{d\psi_n^{(-)}}{dx}|_{0-\epsilon} = \frac{d\psi_n^{(-)}}{dx}|_{0+\varepsilon}, \varepsilon \to 0 \tag{45}$$

$$\frac{d^2\psi_n^{(-)}}{dx^2}|_{0-\epsilon} = \frac{d^2\psi_n^{(-)}}{dx^2}|_{0+\varepsilon} = 0, \varepsilon \to 0 \tag{46}$$

After substituting Eq.(29) into Eq.(27) and performing trivial mathematical calculations, one finds

$$\frac{d^2\alpha_n}{dt^2} + \frac{2}{\tau}\frac{d\alpha_n}{dt} + \omega_{(+),n}^2 \alpha_n = -a(t)\frac{1}{2L}\int_{-L}^{L}dx\psi_n^{(+)} + \frac{1}{m}\int_{-L}^{L}dx\psi_n^{(+)} F_L(x,t) \qquad (47)$$

$$\frac{d^2\beta_n}{dt^2} + \frac{2}{\tau}\frac{d\beta_n}{dt} + \omega_{(-),n}^2 \beta_n = \frac{1}{2L}\tilde{\Gamma}_{zx}\int_{-L}^{L}dxx\psi_n^{(-)} + \frac{1}{m}\int_{-L}^{L}dx\psi_n^{(-)} F_L(x,t) \qquad (48)$$

where $\tau = 2\eta/h$ is the beam's relaxation time and $\omega_{(\pm),n} = 2\pi f_{(\pm),n}$ are its free-vibration angular resonant frequencies[29]

$$f_{(\pm),n} = \frac{1}{2\pi}\sqrt{\frac{EI}{\eta}}k_{(\pm),n}^2 \qquad (49)$$

The eigenvalues $k_{(\pm),n}$ are unambiguously determined from Eq.(34) which is manifestly invariant of the choice of normalisation of modal eigenfunctions

$$\psi_n^{(\pm)}(x) \to N\psi_n^{(\pm)}(x) \qquad (50)$$

The exact solutions of the Eq.(33) for the free-hinged-hinged-hinged-free beam are presented in Supplementary Materials section of this article. The first four force-free eigenmodes of the beam under consideration are depicted in Fig.6 below for $L=0.15$ metres and the locations of side hinges $l = \pm(2 - 2^{1/3})L$:

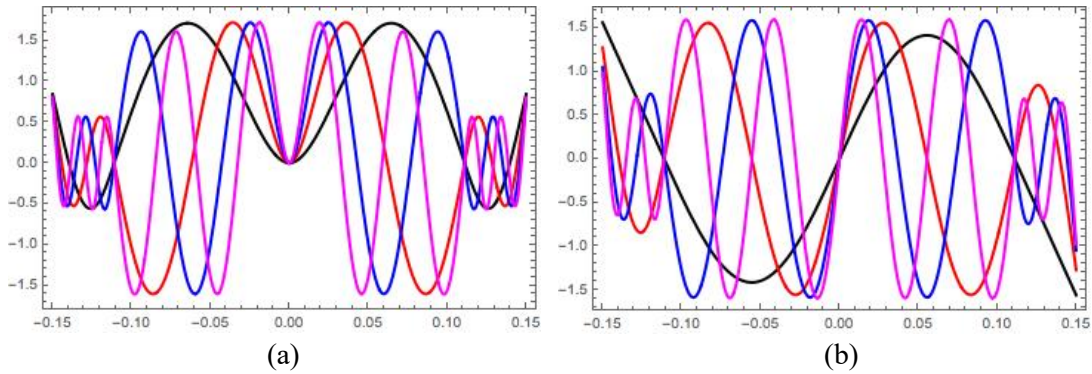

(a)                                     (b)

**Figure 6**. (a) Symmetric eigenmodes of free-hinged-hinged-hinged-free beam (1st - black, 2nd - red, 3rd - blue, 4th – magenta); (b) Mirror-symmetric eigenmodes of free-hinged-hinged-hinged-free beam (1st - black, 2nd - red, 3rd - blue, 4th – magenta).

**Noise Only Driven Free-Hinged-Hinged-Hinged-Free Beam**

The total energy of the free-hinged-hinged-hinged-free beam under consideration consists of a kinetic term and the potential energy stored in the beam's atomic structure[29]

$$W = \frac{\eta}{2}\int_{-L}^{L}dx\left(\frac{\partial Z}{\partial t}\right)^2 + \frac{EI}{2}\int_{-L}^{L}dx\left(\frac{\partial^2 Z}{\partial x^2}\right)^2 \qquad (51)$$

By integrating the potential energy term in Eq.(51) by parts and replacing the beam's transverse displacement $Z(x,t)$ by its modal series expansion in Eq.(29) (the beam's free-ends boundary conditions are taken into account), one finds

$$W = \sum_n (W_{(+),n} + W_{(-),n}) = \frac{m}{2} \sum_n \left[ \left(\frac{d\alpha_n}{dt}\right)^2 + \left(\frac{d\beta_n}{dt}\right)^2 + \omega_{(+),n}^2 \alpha_n^2 + \omega_{(-),n}^2 \beta_n^2 \right] \quad (52)$$

Eq.(52) above is the standard representation of the total energy of multi-mode mechanical oscillator where $\alpha_n$ and $\beta_n$ are its modal mechanical displacement amplitudes. The spectral densities of the latter for the noise only driven beam under consideration are as follows

$$S_{(\pm),n}(\omega) = \frac{D_F}{(\omega^2 - \omega_{(\pm),n}^2)^2 + \left(\frac{2\omega}{\tau}\right)^2} \frac{2L}{m^2} \quad (53)$$

where $D_F$ is the intensity of the Langevin force:

$$\langle F_L(x,t) F_L(x',t') \rangle = D_F \delta(x-x') \delta(t-t') \quad (54)$$

If the beam is in thermal equilibrium with its environment with temperature $T$, the following requirement holds[30]

$$\langle W_{(\pm),n} \rangle = \frac{D_F}{2\pi\eta} \int_0^\infty d\omega \frac{\omega^2 + \omega_{(\pm),n}^2}{(\omega^2 - \omega_{(\pm),n}^2)^2 + \left(\frac{2\omega}{\tau}\right)^2} = k_B T \quad (55)$$

The integration in Eq.(55) yields

$$D_F = 2 k_B T h \quad (56)$$

It is possible now to estimate the thermal fluctuations imposed by the Langevin force upon, say, the mechanical displacement of the beam's free ends under a gravity gradient along its length. In turn, this would allow for an estimate of the thermal limit in measuring gravity gradients by the primary sensing element such as free-hinged-hinged-free beam. One finds from Eq.(30), Eq.(53) and Eq.(56)

$$S(L,\omega) = \sum_n S_{(+),n}(\omega) \psi_n^{(+)2}(L) + \sum_n S_{(-),n}(\omega) \psi_n^{(-)2}(L) \quad (57)$$

where $S(L,\omega)$ is the displacement spectral noise of the beam's free ends. Considering the quasi-static approximation only ($\omega \to 0$), one has

$$S^{1/2}(L,\omega) = \sqrt{\frac{4 k_B T}{m\tau} \left( \sum_n \frac{\psi_n^{(+)2}(L)}{\omega_{(+),n}^4} + \sum_n \frac{\psi_n^{(-)2}(L)}{\omega_{(-),n}^4} \right)} \quad \frac{metres}{\sqrt{Hz}} \quad (58)$$

By combining Eq.(58), Eq.(49) and Eq.(25), one finds the equivalent gravity gradient spectral noise

$$S_\Gamma^{1/2}(L,\omega) = \frac{10^{12}}{L^5} \sqrt{\frac{k_B T}{m\tau} \left( \sum_n \frac{\psi_n^{(+)2}(L)}{k_{(+),n}^8} + \sum_n \frac{\psi_n^{(-)2}(L)}{k_{(-),n}^8} \right)} \quad \frac{Eotvos}{\sqrt{Hz}} \quad (59)$$

It is worth noting that either Eq.(58) or Eq.(59) are invariant of the eigenfunction normalisation factor in Eq.(50) (chosen to be $\sqrt{2L}$) as in the dynamic analysis the total mass of the beam is defined as

$$m = \eta \int_{-L}^{L} dx\, \psi_n^{(\pm)}(x)\psi_n^{(\pm)}(x) = 2L\eta \qquad (60)$$

For the over-constrained free-hinged-hinged-hinged-free beam under consideration only the first ($n=1$) mirror-symmetric eigenmode $\psi_n^{(-)}$ gives the major contribution to the right sides of Eq.(59) and Eq.(60). In case differential displacement measurements are used to measure relative motion of the beam's free ends ($Z(L)-Z(-L)$), then 3 dB noise reduction should be applied to Eq.(59). Tab.1 below gives a few examples for particular beam's dimensions, materials and the locations of side hinges $l = \pm(2 - 2^{1/3})L$.

| Material | Phosphor Bronze | Cu94 Sn6 | | Tungsten Copper | W80 Cu20 | |
|---|---|---|---|---|---|---|
| Length (2L, metres) | 0.3 | 0.35 | 0.4 | 0.3 | 0.35 | 0.4 |
| Width (b, metres) | 0.01 | 0.01 | 0.01 | 0.01 | 0.01 | 0.01 |
| Thickness (d, metres) | 0.00025 | 0.00025 | 0.0003 | 0.00025 | 0.00025 | 0.0003 |
| Mass (m, kg) | 0.0067 | 0.0078 | 0.011 | 0.011 | 0.013 | 0.018 |
| Thermally activated gravity gradient noise (E/√Hz, T=300K, τ=1sec) | 24 | 19 | 14 | 18 | 15 | 11 |
| Free ends' differential displacement under 10 Eotvos gravity gradient (metres) | $4.6 \times 10^{-14}$ | $9.6 \times 10^{-14}$ | $1.3 \times 10^{-13}$ | $3.1 \times 10^{-14}$ | $6.4 \times 10^{-14}$ | $8.6 \times 10^{-14}$ |
| $k_{(-),1}$ (1/metres) / $f_{(-),1}$ (Hz) | 26.2 / 28.4 | 22.5 / 20.9 | 19.7 / 19.2 | 26.2 / 35.3 | 22.5 / 25.9 | 19.7 / 23.8 |
| $k_{(-),2}$ (1/metres) / $f_{(-),2}$ (Hz) | 66.7 / 184.5 | 57.3 / 135.5 | 50.1 / 124.5 | 66.7 / 229.3 | 57.3 / 168.4 | 50.1 / 154.8 |
| $k_{(+),1}$ (1/metres) / $f_{(+),1}$ (Hz) | 46.9 / 90.8 | 40.2 / 66.7 | 35.2 / 61.3 | 46.9 / 112.8 | 40.2 / 82.9 | 35.2 / 76.1 |
| $k_{(+),2}$ (1/metres) / $f_{(+),2}$ (Hz) | 82.6 / 281.8 | 70.8 / 207.1 | 62 / 190.2 | 82.6 / 350.2 | 70.8 / 257.3 | 62 / 236.4 |
| $\psi_1^{(-)}(L)$ / $\psi_1^{(+)}(L)$ | 1.56 / 0.84 | 1.56 / 0.84 | 1.56 / 0.84 | 1.56 / 0.84 | 1.56 / 0.84 | 1.56 / 0.84 |
| $\psi_2^{(-)}(L)$ / $\psi_2^{(+)}(L)$ | 1.27 / 0.8 | 1.27 / 0.8 | 1.27 / 0.8 | 1.27 / 0.8 | 1.27 / 0.8 | 1.27 / 0.8 |

**Table 1**. Numerical parameters of free-hinged-hinged-hinged-beam with different lengths and cross sections for two different materials (Phosphor Bronze and Tungsten Copper alloys); calculated the first two modal eigenvalues and mechanical resonant frequencies for both mirror-symmetric and symmetric eigenfunctions; calculated differential mechanical displacement of the beam's free ends under 10 Eotvos gravity gradient and the thermally activated equivalent gravity gradient noise.

**Concluding Remarks**

The static and dynamic analysis of the free-hinged-hinged-hinged-free beam is the first step in presenting a novel design of a primary sensing element having a single sensitivity axis that can be used in future advanced gravity gradiometers. The latter should be capable of operating in any orientation with respect to the Earth's acceleration of gravity and in a limited space that is typical for such unmanned mobile platforms as UAVs (either airborne or submersible) and drones. From Tab.1 above it follows that the beam's length is the most critical factor for either increasing its sensitivity to a gravity gradient along its length or reducing the thermally activated gravity gradient noise. The latter is the limiting factor for this type of gravity gradiometer. A median baseline (2L) of 0.35 metres, based on the free-hinged-hinged-hinged-free beam and made of a Tungsten/Copper composite alloy, would provide better than 10 Eotvos resolution at 3 seconds measurement time. The corresponding mechanical displacement measurements at the level of ~ $10^{-14}$ metres/$\sqrt{Hz}$ can be provided by a number of currently available room-temperature techniques (capacitive, microwave, optical) including new emerging techniques such as quantum and quantum enabled sensing[31]. Further improvement could be provided by applying a modulation-demodulation technique that can squeeze the effective measurement time below the beam's relaxation time (this approach is currently under development and will be discussed elsewhere). The latter approach is preferred as it also allows for effective removal of 1/f noise (zero-point drift) caused by temperature instabilities, mechanical creep in the primary sensing element, voltage drift in operational amplifiers and other electronic components. The most important feature of the distributed sensing element as the free-hinged-hinged-hinged-free beam is that it allows for simultaneous measurement of the dynamic displacements of the beam's different spatial locations representing true standing waves. If combined in a proper manner, this could cancel out the effect of one of the most disturbing factors for gravity gradiometers, if being deployed on moving platforms, namely large kinematic and uniform gravitational accelerations applied to the primary sensing element and its frame of reference. This approach is entirely new in gravity gradiometry and may open the door to the most desired use of gravity gradiometers in the strapped-down mode onboard commercial drones and other unmanned platforms.

**Acknowledgements**

This study was financially supported by Lockheed Martin Corporation (USA). One of the authors (AV) is thankful to Prof David Blair, Dr John Winterflood (ARC Centre of Excellence for Gravitational Wave Discovery, UWA) and Prof Michael Tobar (Quantum Technology and Dark Matter Research Laboratory, UWA ) for reading the manuscript and providing useful comments and suggestions.

**Competing interests**

The authors declare no competing interests.

**Supplementary Materials**

The modal mirror-symmetric eigenfunctions for the free-hinged-hinged-hinged-free beam:

$l \leq x \leq L$

$$\psi_n^{(-)}(x) = \frac{(-1)^n n\pi (L-l)\left(\text{Cosh}[\xi_{(-),n}]\text{Sin}\left[\xi_{(-),n}\frac{x-l}{L-l}\right] + \text{Cos}[\xi_{(-),n}]\text{Sinh}\left[\xi_{(-),n}\frac{x-l}{L-l}\right]\right)}{\xi_{(-),n} l (\text{Cos}[\xi_{(-),n}] + \text{Cosh}[\xi_{(-),n}])}$$

$-l \leq x \leq l$

$$\psi_n^{(-)}(x) = \text{Sin}[\pi n(x/l)]$$

$-L \leq x \leq -l$

$$\psi_n^{(-)}(x) = \frac{(-1)^n n\pi (L-l)\left(\text{Cosh}[\xi_{(-),n}]\text{Sin}\left[\xi_{(-),n}\frac{x+l}{L-l}\right] + \text{Cos}[\xi_{(-),n}]\text{Sinh}\left[\xi_{(-),n}\frac{x+l}{L-l}\right]\right)}{\xi_{(-),n} l (\text{Cos}[\xi_{(-),n}] + \text{Cosh}[\xi_{(-),n}])}$$

where $\xi_{(-),n}$ approximates as follows

$$\xi_{(-),n} \cong \frac{10^{-8}}{n^8} + \text{Tanh}[1.5778901\pi(n-1)]\left(\frac{\pi}{4} + \pi(n-1)\right)$$

$n = 1,2,3 \ldots$

The modal symmetric eigenfunctions for the free-hinged-hinged-hinged-free beam:

$l \leq x \leq L$

$$\psi_n^{(+)}(x) = \frac{(L-l)(1 - \text{Csch}[\xi_{(+),n}]\text{Sin}[\xi_{(+),n}])(\text{Sec}[\xi_{(+),n}]\text{Sin}[\xi_{(+),n}\frac{x-l}{L-l}] + \text{Sech}[\xi_{(+),n}]\text{Sinh}[\xi_{(+),n}\frac{x-l}{L-l}])}{l(1 + \text{Cos}[\xi_{(+),n}]\text{Sech}[\xi_{(+),n}])}$$

$0 \leq x \leq l$

$$\psi_n^{(+)}(x) = \left(\text{Cosh}[\xi_{(+),n}\frac{x}{l}] - \text{Coth}[\xi_{(+),n}]\text{Sinh}[\xi_{(+),n}\frac{x}{l}]\right)\text{Tan}[\xi_{(+),n}] - \text{Sec}[\xi_{(+),n}]\text{Sin}[\xi_{(+),n}\frac{l-x}{l}]$$

$-l \leq x \leq 0$

$$\psi_n^{(+)}(x) = \left(\text{Cosh}\left[\xi_{(+),n}\frac{x}{l}\right] + \text{Coth}[\xi_{(+),n}]\text{Sinh}[\xi_{(+),n}\frac{x}{l}]\right)\text{Tan}[\xi_{(+),n}] - \text{Sec}[\xi_{(+),n}]\text{Sin}[\xi_{(+),n}\frac{l+x}{l}]$$

$-L \leq x \leq -l$

$$\psi_n^{(+)}(x) = -\frac{(L-l)(1 - \text{Csch}[\xi_{(+),n}]\text{Sin}[\xi_{(+),n}])(\text{Sec}[\xi_{(+),n}]\text{Sin}[\xi_{(+),n}\frac{x+l}{L-l}] + \text{Sech}[\xi_{(+),n}]\text{Sinh}[\xi_{(+),n}\frac{x+l}{L-l}])}{l(1 + \text{Cos}[\xi_{(+),n}]\text{Sech}[\xi_{(+),n}])}$$

where $\xi_{(+),n}$ approximates as follows

$$\xi_{(+),n} = \text{Tanh}[1.5778901\pi n](\frac{\pi}{4} + \pi n)$$

$n = 1,2,3 \ldots$

All modal eigenfunctions above can be arbitrary normalised by the following transfer

$$\psi_n^{(\pm)}(x) \to \frac{N}{\sqrt{\int_{-L}^{L} dx \, \psi_n^{(\pm)}(x)\psi_n^{(\pm)}(x)}} \psi_n^{(\pm)}(x)$$

where $N$ is a normalisation factor (chosen to be $\sqrt{2L}$ throughout this paper).

The modal eigenvalues $k_{(\pm),n}$ for the whole beam's length are found from Eq.34 as follows

$$k_{(\pm),n} = \left( \frac{\int_{-L}^{L} dx \psi_n^{(\pm)}(x) \frac{d^4}{dx^4} \psi_n^{(\pm)}(x)}{\int_{-L}^{L} dx \, \psi_n^{(\pm)}(x)\psi_n^{(\pm)}(x)} \right)^{1/4}$$